\documentclass[11pt,twoside]{article}
\usepackage{asp2010}

\resetcounters

\begin{document}

\title{Long period sdB + MS binaries with Mercator.}
\author{Joris Vos$^1$, Roy H. \O{}stensen$^1$ and Hans Van Winckel$^1$
\affil{$^1$Instituut voor Sterrenkunde, KU Leuven, B-3001 Leuven, Belgium} }

\begin{abstract}
The predicted orbital period distribution of the subdwarf-B (sdB) population is bi-modal with a peak at short ( $<$ 10 days) and long ( $>$ 250 days) periods. Observationally,  many short-period sdB systems are indeed known. The predicted long period peak is missing and orbits have only been determined for a few long-period systems. As these predictions are based on poorly understood binary-interaction processes, it is of prime importance to confront the predictions to well-constrained observational data. We therefore initiated a monitoring program to find and characterize long-period sdB stars with the 1.2\,m Mercator telescope. Here we present the results of this program.
\end{abstract}

\section{Introduction}
Binary interaction plays an important role in the formation of subdwarf-B (sdB) stars \citep{Mengel76}. Currently, there is a consensus that sdB stars are formed by binary evolution only, and several of these evolutionary channels are described. Close binary systems can be formed in a common envelope (CE) ejection channel \citep{Paczynski76}, while stable Roche-lobe overflow (RLOF) can produce wide sdB binaries \citep{Han00, Han02}. An alternative formation channel forming a single sdB star is the double white dwarf (WD) merger, where a pair of white dwarfs spiral in to form a single sdB star \citep{Webbink84}.

\citet{Han02, Han03} addressed these three binary formation mechanisms, and performed binary population synthesis (BPS) studies for two kinds of CE ejection channels, two possible stable RLOF channels and the WD merger channel. The CE ejection channels produce close binaries with periods of $P_{\rm{orb}}$ = 0.1 -- 10 d, while sdB  binaries formed through stable RLOF have orbital periods ranging from 10 to 500 days. An alternative stable RLOF channel based on the $\gamma$-formalism is described by \citet{Nelemans10} and can produce sdB binaries with periods on the order of years. All of these channels predict a very small mass range peaking at $M_{\rm{sdB}}$ = 0.47 $\pm$ 0.05 $M_{\odot}$.

Many observational studies have focused on short-period sdB binaries, and over 100 of these systems are currently known \citep[Appendix A]{Geier11}. These observed short-period sdB binaries correspond very well with the results of BPS studies. However, only few long period sdB binaries are known, and the current studies show that there are still large discrepancies between theory and observations \citep{Geier13}. The binary interaction processes governing the formation of long period sdB binaries are very complex, and several aspects are still not understood.

In 2009 a radial velocity monitoring program of long period sdB + MS binaries was started with HERMES at the Mercator telescope. The original sample including six sdB and two sdO binaries was presented by \citet{Oestensen11, Oestensen12}. Here we present orbital and spectroscopic parameters of these six sdB binaries. The analysis of PG\,1104+243 was discussed in detail in \citet{Vos12}, while BD$+$29$^{\circ}$3070, BD$+$34$^{\circ}$1543 and Feige\,87 are discussed in \citet{Vos13}. In this paper we will elaborate on Balloon\,82800003\footnote{Balloon\,82800003 = TYC\,3871$-$835$-$1, RA: 15:15:38.20 Dec: 56:53:45.00} and BD$-$7$^{\circ}$5977.

\section{Observations}
High resolution spectroscopic observations of all six sdB binaries were obtained with the HERMES spectrograph (High Efficiency and Resolution Mercator Echelle Spectrograph, R = 85\,000, 55 orders, 3770-9000 \AA, \citealt{Raskin11}) attached to the 1.2-m Mercator telescope at the Roque de los Muchachos Observatory, La Palma. HERMES is connected to the Mercator telescope by an optical fiber, and is located in a temperature controlled enclosure to ensure optimal wavelength stability. In \citet[Sect.\,2]{Vos12} the wavelength stability was checked, using 38 radial velocity standard stars of the IAU observed over a time span of 1481 days, and a standard deviation of 80 m~s$^{-1}$ with a non-significant shift to the IAU radial velocity standard scale was found. The number of spectra taken between June 2009 and January 2013 for each object in the sample is given in Table \ref{tb-sample}. HERMES was used in high-resolution mode, and Th-Ar-Ne exposures were made at the beginning and end of the night. The exposure 
time of the science observations was adapted to reach a signal-to-noise ratio (S/N) of 25 in the $V$--band. The HERMES pipeline v5.0 was used for the basic reduction of the spectra.

\begin{table}[!ht]
\caption{Overview of the survey sample}\label{tb-sample}
\smallskip
\begin{center}
{\small
\begin{tabular}{llrrrr}
\tableline
\noalign{\smallskip}
ID	&	Target name	&	Spectral Class	&	$m_V$	&	Total Spectra 	&	Total Photometry\\
\noalign{\smallskip}
\tableline
\noalign{\smallskip}
1	&	PG\,1104$+$243		&	sdB + G0	&	11.3	&	38	&	11	\\
2	&	BD$+$29$^{\circ}$3070	&	sdB + F8	&	10.4	&	31	&	17	\\
3	&	BD$+$34$^{\circ}$1543	&	sdB + F8	&	10.1	&	30	&	14	\\
4	&	Feige\,87		&	sdB + G2	&	11.7	&	33	&	11	\\
5	&	Bal\,82800003		&	sdB + G0	&	11.4	&	29	&	9	\\
6	&	BD$-$7$^{\circ}$5977	&	sdB + K2III	&	10.5	&	29	&	9	\\
\noalign{\smallskip}
\tableline
\end{tabular}
}
\end{center}
\end{table}

\section{Radial velocity analysis}
The radial velocities of the MS components are determined with the cross-correlation (CC) method of the HERMES pipeline (\textsc{hermesvr}), based on a discrete number of line positions. This is possible because the sdB component has only a few H and He lines, which are avoided in the CC. For all MS components a G2-type mask was used on orders 55-74 (4780 - 6530\,\AA) as these orders give the best compromise between maximum S/N for G-K type stars and absence of telluric influence.

As only very few spectral lines of the sdB component are visible in the composite spectra a different technique is nessessary to obtain its radial velocities. The only spectral line that is not contaminated by the MS components is the He\,I blend at 5875\,\AA. To derive radial velocities based on only one blended line, a cross correlation with a high-resolution synthetic sdB spectrum from the LTE grids of \citet{Heber00} was used. The cross correlation (CC) is carried out in wavelength space, and the error is determined by performing a Monte-Carlo (MC) simulation in which Gaussian noise is added to the observed spectra after which the CC is repeated.
A more elaborate explanation of the radial velocity determination is given in \citet{Vos12, Vos13}.

The orbital parameters of the sdB and MS components are calculated by fitting a Keplerian orbit to the radial velocity measurements, while adjusting the period ($P$), time of periastron ($T_0$), eccentricity ($e$), angle of periastron ($\omega$), two amplitudes ($K_{\rm{MS}}$ and $K_{\rm{sdB}}$) and two systemic velocities ($\gamma_{\rm{MS}}$ and $\gamma_{\rm{sdB}}$). As a first guess for these parameters, the results of \citet{Oestensen11} were used. The radial velocity measurements were weighted according to their errors as $w = 1/\sigma$. For each system, the \citet{Lucy71} test was used to check if the orbit is significantly eccentric. In the fitting process, the system velocities of both components are allowed to vary independently of each other, to allow for gravitational redshift effects in the sdB component (see \citet[Sect. 4]{Vos12}). The uncertainties on the final parameters are obtained using 5000 iterations in a Monte-Carlo simulation where the radial velocities were perturbed based on their 
errors. 

\begin{figure}[!t]
\centering
\includegraphics{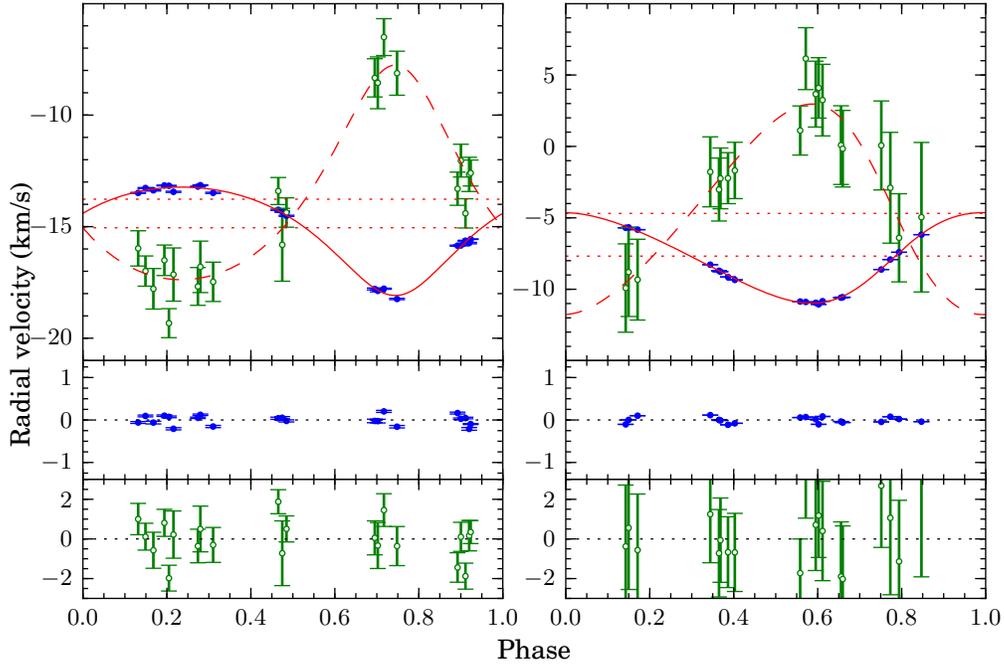}
\caption{The radial velocity curves for Bal\,82800003 (left) and BD$-$7$^{\circ}$5977 (right). Top: spectroscopic orbital solution (solid line: MS, dashed line: sdB), and the observed radial velocities (filled blue circles: MS component, open green circles: sdB component). The measured system velocities of both components are shown by a dotted line. Middle: residuals of the MS component. Bottom: residuals of the sdB component.}\label{fig-rvcurves}
\end{figure}

For Bal\,82800003 we find a period of 1363 $\pm$ 25 days, while for BD$-$7$^{\circ}$5977 a period of 1195 $\pm$ 30 days is measured. Both systems are clearly eccentric with an eccentricity of respecivily 0.25 $\pm$ 0.02 and 0.16 $\pm$ 0.01. The radial velocity curves together with the orbital solutions are plotted in Fig. \ref{fig-rvcurves}. The spectroscopic parameters of all six sdB binaries are summarized in Table \ref{tb-specpar}.

\begin{table}[!ht]
\caption{Summary of the spectroscopic orbital solutions for the MS and sdB component of all six sdB + MS binaries.}\label{tb-specpar}
\smallskip
\begin{center}
{\small
\begin{tabular}{lcccccc}
\tableline
\noalign{\smallskip}
ID	&	P	&	e	&	q	&	$K_{\rm{MS}}$	&	$K_{\rm{sdB}}$	&	$\gamma$	\\
	&	(d)	&		&		&	(km s$^{-1}$)	&	(km s$^{-1}$)	&	(km s$^{-1}$)	\\
\noalign{\smallskip}
\tableline
\noalign{\smallskip}
1	&	753 $\pm$ 3	&	$<$ 0.002	&	0.64 $\pm$ 0.01	&	4.42 $\pm$ 0.08	&	6.9 $\pm$ 0.2	&	-15.17 $\pm$ 0.07	\\
2	&	1283 $\pm$ 63	&	0.15 $\pm$ 0.01	&	0.39 $\pm$ 0.01	&	6.53 $\pm$ 0.30	&	16.6 $\pm$ 0.6	&	-57.58 $\pm$ 0.36	\\
3	&	972 $\pm$ 2	&	0.16 $\pm$ 0.01	&	0.57 $\pm$ 0.01	&	5.91 $\pm$ 0.07	&	10.3 $\pm$ 0.2	&	32.10 $\pm$ 0.06	\\
4	&	936 $\pm$ 2	&	0.11 $\pm$ 0.01	&	0.55 $\pm$ 0.01	&	8.19 $\pm$ 0.11	&	15.0 $\pm$ 0.2	&	32.98 $\pm$ 0.08	\\
5	&	1363 $\pm$ 25	&	0.25 $\pm$ 0.02	&	0.51 $\pm$ 0.02	&	2.43 $\pm$ 0.03	&	4.8 $\pm$ 0.3	&	-15.06 $\pm$ 0.03	\\
6	&	1195 $\pm$ 30	&	0.16 $\pm$ 0.01	&	0.43 $\pm$ 0.01	&	2.77 $\pm$ 0.02	&	6.1 $\pm$ 0.8	&	-8.73 $\pm$ 0.02	\\
\noalign{\smallskip}
\tableline
\end{tabular}
}
\end{center}
\end{table}

\section{Spectral Energy Distribution analysis}
The spectral-energy distribution (SED) of the systems can be used to determine the spectral type of the MS and sdB component. We used photometric SEDs which were fitted with model SEDs to determine both the effective temperature and surface gravity of both components. To collect the photometry of both systems the subdwarf database\footnote{http://catserver.ing.iac.es/sddb/} \citep{Oestensen06}, which contains a compilation of data on hot subdwarf stars collected from the literature, is used. These photometric measurements are supplemented with photometry obtained from several other catalogs.

The SED fitting method used here is described in \citet{Vos12, Vos13}. The observed photometry is fitted with a synthetic SED integrated from model atmospheres, using a combination of a grid based fitting approach followed by a least-squares minimizer. For the MS component Kurucz atmosphere models \citep{Kurucz93} ranging in effective temperature from 4000 to 9000 K, and in surface gravity from $\log{g}$=3.0 dex (cgs) to 5.0 dex (cgs) are used. For the hot sdB component TMAP (T\"{u}bingen NLTE Model-Atmosphere Package, \citealt{Werner03}) atmosphere models with a temperature range from 20000 K to 50000 K, and $\log{g}$ from 4.5 dex (cgs) to 6.5 dex (cgs) are used.

Using the nine photometric measurements that are available for Bal\,82800003 we find that this binary consists of a rather cold sdB component with $T_{\rm{eff, sdB}}$ = 22500 $\pm$ 3500 K and $\log{g}_{\rm{sdB}}$ = 5.12 $\pm$ 0.4 dex, and a MS companion with $T_{\rm{eff, MS}}$ = 5970 $\pm$ 350 K and $\log{g}_{\rm{MS}}$ = 4.10 $\pm$ 0.35 dex. The system has a reddening of E($B-V$) = 0.01$_{-0.01}^{+0.025}$ mag, which is consistent with the maximum reddening E($B-V$)$_{\rm{max}}$ = 0.035 derived from the dust maps of \citet{Schlegel98}. The probability distibution of the MS companion follows a Gaussian pattern, but for the sdB companion there is a much larger spread, indicating that the derived effective temperature and surface gravity are not as reliable.

The SED of BD$-$7$^{\circ}$5977 is clearly different from that of all other binaries in the sample, as it is dominated by the cool companion, indicating that the companion is more luminous than a main sequence star. The sdB component is only visible in the UV. The SED fit results in: $T_{\rm{eff, sdB}}$ = 29000 $\pm$ 400 K, $\log{g}_{\rm{sdB}}$ = 5.02 $\pm$ 0.4 dex, $T_{\rm{eff, MS}}$ = 4720 $\pm$ 250 K and $\log{g}_{\rm{MS}}$ = 2.86 $\pm$ 0.35 dex. The reddening is rather high at E($B-V$) = 0.034 $\pm$ 0.034 mag.

\begin{figure}[!t]
\centering
\includegraphics{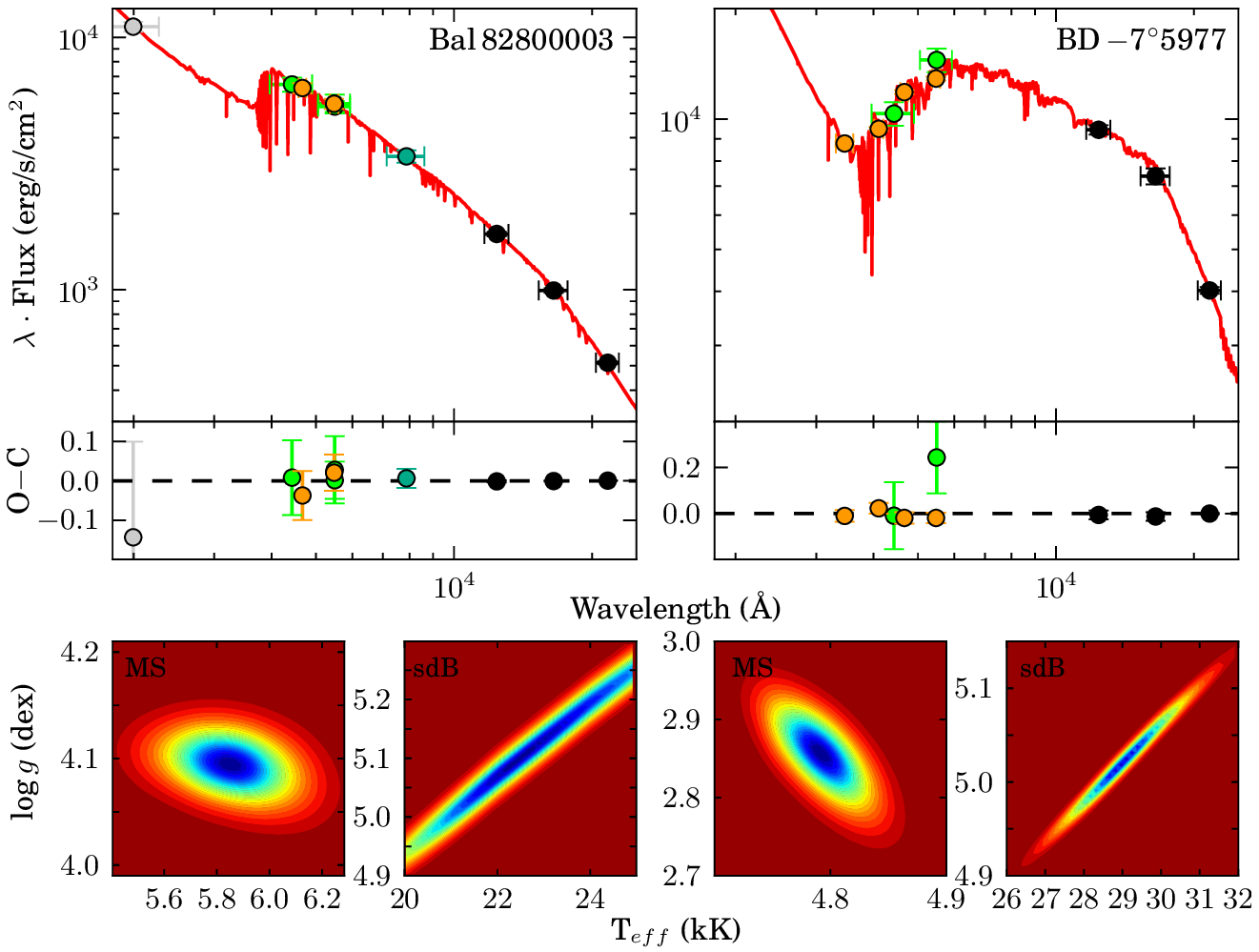}
\caption{}\label{fig-sedfit}
\end{figure}

\begin{table}[!ht]
\caption{Photometric parameters of the systems.}\label{tb-photpar}
\smallskip
\begin{center}
{\small
\begin{tabular}{lccccc}
\tableline
\noalign{\smallskip}
ID	&	T$_{\rm{eff,MS}}$	&	$\log g_{\rm{MS}}$	&	T$_{\rm{eff,sdB}}$	&	$\log g_{\rm{sdB}}$	&	E($b-v$)	\\
	&	(K)	&	(dex)	&	(K)	&	(dex)	&	(mag)	\\
\noalign{\smallskip}
\tableline
\noalign{\smallskip}
1	&	5930 $\pm$ 160	&	4.29 $\pm$ 0.05	&	33500 $\pm$ 1200	&	5.85 $\pm$ 0.08	&	0.001 $\pm$ 0.017	\\
2	&	6180 $\pm$ 420	&	4.32 $\pm$ 0.50	&	25900 $\pm$ 3000	&	5.56 $\pm$ 0.39	&	0.010 $\pm$ 0.030	\\
3	&	6150 $\pm$ 250	&	4.07 $\pm$ 0.28	&	36900 $\pm$ 2100	&	5.79 $\pm$ 0.20	&	0.006 $\pm$ 0.030	\\
4	&	5980 $\pm$ 325	&	4.36 $\pm$ 0.42	&	27300 $\pm$ 2700	&	5.55 $\pm$ 0.40	&	0.006 $\pm$ 0.030	\\
5	&	5970 $\pm$ 350	&	4.10 $\pm$ 0.35	&	22500 $\pm$ 3500	&	5.12 $\pm$ 0.40	&	0.010 $\pm$ 0.025	\\
6	&	4720 $\pm$ 250	&	2.86 $\pm$ 0.35	&	29000 $\pm$ 4000	&	5.02 $\pm$ 0.40	&	0.034 $\pm$ 0.034	\\
\noalign{\smallskip}
\tableline
\end{tabular}
}
\end{center}
\end{table}

\section{Conclusions}
Long time base spectroscopy from the Mercator telescope was used to solve the orbits of six sdB binaries, and resulted in orbital periods of 750 to over 1300 days. Furthermore, five out of six systems have a significant eccentric orbit instead of the predicted circular orbits. The sdB components of all six systems are consistent with a canonical post-core-helium-flash model with a mass around 0.47 $M_{\odot}$. During the red giant phase of the sdB progenitor the tidal forces between the sdB progenitor and its companion should circularize the orbit very efficiently \citep{Zahn77}, and the further evolution provides very few possibilities to re-introduce eccentricity to the orbit. 

\citet{Deca12} proposed the hierarchial triple merger scenario of \citet{Clausen11} for the possibly eccentric sdB + K system PG\,1018-047, where the K-type companion would not have been involved in the evolution of the sdB component. However, such a scenario seems too unlikely to be observed frequently.
Another possible explanation for these eccentric systems would be that tidal circularization during the RGB and the RLOF phase is not as efficient as currently assumed. In highly eccentric systems the mass transfer through RLOF and mass loss due to stellar winds will not be constant over the orbit. \citet{Bonacic08} studied this effect in binary systems with an AGB star, and found that tidally enhanced mass loss from the AGB star at orbital phases closer to periastron can work efficiently against the tidal circularization of the orbit. Further theoretical studies could investigate if this eccentricity enhancement mechanism can work in in the progenitors of sdB+MS binaries as well.

\acknowledgements 
Based on observations made with the Mercator Telescope, operated on the island of La Palma by the Flemish Community, at the Spanish Observatorio del Roque de los Muchachos of the Instituto de Astrofísica de Canarias.
Based on observations obtained with the HERMES spectrograph, which is supported by the Fund for Scientific Research of Flanders (FWO), Belgium , the Research Council of K.U.Leuven, Belgium, the Fonds National Recherches Scientific (FNRS), Belgium, the Royal Observatory of Belgium, the Observatoire de Genève, Switzerland and the Thüringer Landessternwarte Tautenburg, Germany.
The research leading to these results has received funding from the European Research Council under the European Community's Seventh Framework Programme (FP7/2007--2013)/ERC grant agreement N$^{\underline{\mathrm o}}$\,227224 ({\sc prosperity}), as well as from the Research Council of K.U.Leuven grant agreements GOA/2008/04 and GOA/2013/012.
The following Internet-based resources were used in research for this paper: the NASA Astrophysics Data System; the SIMBAD database and the VizieR service operated by CDS, Strasbourg, France; the ar$\chi$ive scientific paper preprint service operated by Cornell University.
This publication makes use of data products from the Two Micron All Sky Survey, which is a joint project of the University of Massachusetts and the Infrared Processing and Analysis Center/California Institute of Technology, funded by the National Aeronautics and Space Administration and the National Science Foundation.

\bibliography{bibliography}

\end{document}